\documentclass[aps,pre,reprint,amsmath,amssymb]{revtex4-1}

\usepackage{graphicx,subfigure}
\usepackage{dcolumn}
\usepackage[usenames]{xcolor}
\usepackage{color}
\usepackage{siunitx}
\usepackage{bm}
\usepackage{upgreek}
\usepackage{ulem}

\makeatletter
\def\btt#1{\texttt{\@backslashchar#1}}%
\DeclareRobustCommand\bblash{\btt{\@backslashchar}}%
\makeatother

\begin{document}

\title{A novel method for measuring electric field induced dipole moments of metal-dielectric Janus particles in nematic liquid crystals}
\date{\today}
\author{Dinesh Kumar Sahu and Surajit Dhara}
\email{sdsp@uohyd.ernet.in} 
\affiliation{School of Physics, University of Hyderabad, Hyderabad-500046, India}

\begin{abstract}

Janus particles are special types of nano or microparticles possessing at least two surfaces with distinct physical or chemical properties. The most studied Janus particles are the metal-dielectric particles, in which half surface of dielectric particles is coated with a very  thin layer of metals. The external electric field induces dipole moment, and consequently the particles exhibit self-assembled dynamic structures in concentrated aqueous suspensions. Here, we study metal-dielectric Janus particles in a nematic liquid crystal under AC electric field and demonstrate a novel method for measuring effective induced dipole moments of the particles, through competition between elastic and electrostatic (Coulomb) forces of the two particles. The calculated polarisability of the particles based on a simple model agrees well with the effective polarisability measured in the experiments. Our findings have important bearing on functional materials based on metal-dielectric Janus particles dispersed in an anisotropic medium.

\end{abstract}
\preprint{HEP/123-qed}
\maketitle

\section{Introduction}

In recent years, Janus particles have emerged as a new class of colloids that finds potential applications in interdisciplinary areas ranging from physics to biology, medicine and chemistry. For metal-dielectric Janus particles, the asymmetric surfaces behave differently under external electric field and the resulting self-assembly as well as dynamics are markedly different than regular colloids with symmetric surface~\cite{1,2}. Janus particles are used as building blocks for reconfigurable colloidal structures and superstructures on different length scales, which are potential for designing new and functional materials with complex architecture~\cite{stv1}. Apart from applications, these particles are also important in studying nonequilibrium phenomena in soft matter physics.  The imposed electric field induces dipole moment, whose magnitude depends on the electrical properties of the suspending medium and plays a crucial role in determining their equilibrium structures and dynamics. Although several experimental studies have been reported, the magnitude of the induced dipole moments of such particles has not been measured experimentally so far in any dispersive medium. Here, we demonstrate a novel method of measuring the effective dipole moments of the Janus particles in a thermotropic nematic liquid crystal. 

\begin{figure}[!ht]
\center\includegraphics[scale=0.43]{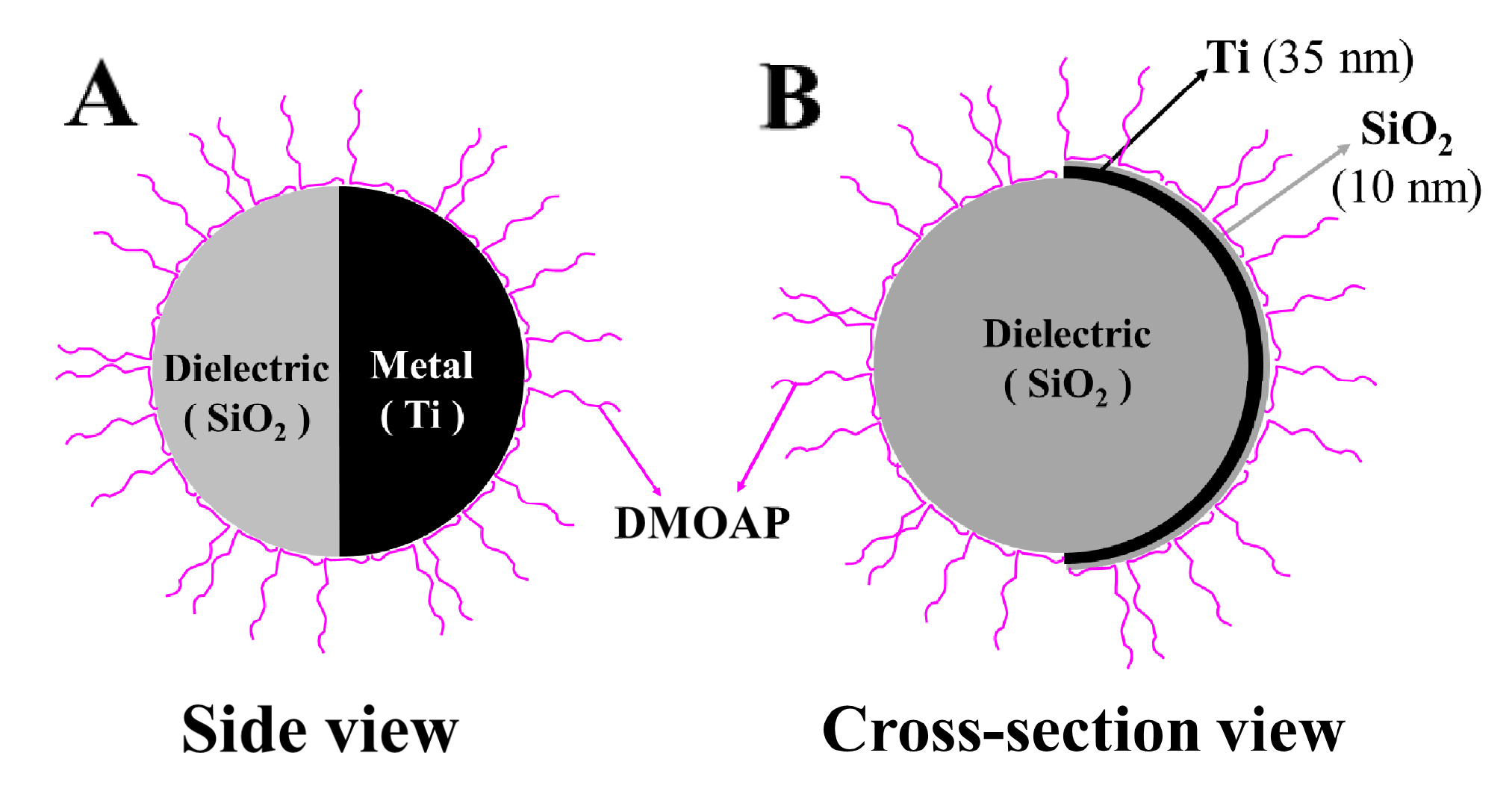}
\caption{(A) Top and (B) cross-section views of a metal-dielectric Janus particle prepared by coating 35 nm Ti layer on SiO$_2$ microparticle. On top of Ti, 10 nm of SiO$_2$ is coated. Magenta colour represents DMOAP molecules, anchored to the surface.}
\label{fig:figure1}
\end{figure}
 
When  microparticles with normal anchoring condition for the director ${\bf \hat{n}}$ (the average direction of molecular orientation) are dispersed in a uniformly aligned nematic liquid crystal (NLC), the particles induce ring~\cite{ram,abot} or point~\cite{lub,stark} defects, resulting, respectively, in quadrupolar or dipolar elastic distortions. Unlike conventional colloids in aqueous solutions, these particles experience structural forces, which arise due to the elastic distortions and local variations of orientational order of the liquid crystals and  show fascinating self-assembled colloidal structure~\cite{rev,igor1,ivan1,igor,ivan2,ivan3}. The interaction potential between two elastic dipoles and quadrupoles are  given by $U_{DD}(r,\theta)=C_{1}K(1-3\cos^{3}\theta)/r^{3}$ and $U_{QQ}(r,\theta)=C_{2}K(3-30\cos^{2}\theta+35\cos^{4}\theta)/r^{5}$, respectively, where $C_{1}$, $C_{2}$ are constants, $K$ is the average elastic constant of the medium, $r$ is the separation between them and $\theta$ is the angle between the vector connecting the centers of particles and the far-field orientation of the nematic director  ~\cite{stark}. Liquid crystals are weak electrolytes and the external electric field drives the ions in opposite direction of the field and generates electroosmotic flows surrounding the particles due to the nonlinear electrophoresis~\cite{oleg1,oleg2}. Under imposed electric field, the dipolar particles become motile along the nematic director as the fore-aft (left-right) symmetry of the electroosmotic flows  is broken, whereas the quadrupolar particles by contrast exhibit no motility due to the fore-aft symmetery ~\cite{oleg1,oleg2,oleg3}. When metal-dielectric Janus particles are suspended in NLCs, from the mechanics point of view they are either elastic dipoles or quadrupoles but display Janus character under the application of external electric field. Recently it has been shown that the fore-aft symmetry of quadrupolar Janus particles is broken and the particles pick up a direction of motion (above a certain electric field) depending on the orientation of the Janus vector $\bf{\hat{s}}$ (normal to the metal-dielectric interface) with respect to the director~\cite{sd}.

Here, we study the effect of small amplitude AC electric fields on a static pair of assembled quadrupolar Janus particles. The particles experience dipolar Coulomb as well as elastic forces of the medium which are directed oppositely along the line joining their centres. We demonstrate a competing effect of the two forces, which allows us to measure the effective induced dipole moments of the particles. The applicability of the technique is limited to liquid crystals but relevant to all microscopic particles, irrespective of shapes.

\section{Experimental}

Metal-dielectric Janus particles are prepared using directional deposition of metal onto dry silica particles (SiO$_{2}$) of diameter $2a = 3.0 \pm 0.2$ $\upmu$m (Bangs Laboratories, USA) in vacuum~\cite{stv1}. Approximately $2 \% $ suspension ($25$ $\upmu$l) of silica particles is spread on a half glass side, which is pretreated with Piranha solution and dried to form monolayer. Next, a thin Titanium (Ti) layer of thickness $35$ nm is deposited vertically using electron-beam deposition. On top of this, a thin layer of SiO$_{2}$ film (10-15 nm) is deposited. Then the slides with silica monolayer are washed thoroughly with deionized water (DI) and isopropyl alcohol. The particles are detached from the slides by ultrasonication in deionized water for 20 s. Further, the collected particles are sonicated for 5 minutes to breakup any agglomeration of the Janus particles. The surface of the Janus particles is coated with N, N-dimetyl-N-octadecyl-$3$  aminopropyl-trimethoxysilyl chloride (DMOAP) in order to induce perpendicular (homeotropic) orientation of the liquid crystal director~\cite{zu1,igor2,igor3}.  Schematic representation of side and cross-section views of a Janus particle with successive coatings are shown in Figs.\ref{fig:figure1}A and 1B, respectively. A small quantity (0.001 wt$\%$) of DMOAP coated Janus particles is dispersed in nematic liquid crystal, MLC-6608 (Merck). It exhibits the following phase transitions: SmA $-30^{\circ}$C N $90^{\circ}$C Iso. The dielectric anisotropy of MLC-6608 is negative ($ \Delta\epsilon = \epsilon_{\parallel} - \epsilon_{\perp} = -3.3$, where $\epsilon_{||}$ and $\epsilon_{\perp}$ are the dielectric permittivities for electric field {\bf E}, parallel and perpendicular to ${\bf\hat n}$) whereas the conductivity anisotropy is positive ($ \Delta\sigma =\sigma_{\parallel}-\sigma_{\perp}$ $\simeq$ $6\times10^{-10}$ Sm\textsuperscript{-1} at $100$ Hz) (Supplemental Material~\cite{sup}). No electroconvection was observed in the experimental field and frequency range.\\

 The experimental cells are prepared using two indium-tin-oxide (ITO) coated glass plates. The plates are spin coated with polyimide AL-$1254$ (JSR Corporation, Japan) and cured at $180^{0}$C for $1$ hour. They are then rubbed unidirectionally using a bench top rubbing machine (HO-IAD-BTR-$01$) for homogeneous or planar alignment of the nematic director ${\bf\hat{n}}$. The ITO plates are separated by spherical spacers of diameter $5.0$ $\upmu$m, making the rubbing directions antiparallel, and then sealed with UV-curing optical adhesive (Norland, NOA-$81$). Schematic diagram of a cell is shown in Fig.\ref{fig:figure2}B. The output of a function generator (Tektronix, AFG $3102$) is connected to a voltage amplifier for applying sinusoidal voltage to the cell (Fig.\ref{fig:figure2}C). An inverted polarising optical microscope (Nikon Ti-U) with water immersion objective (Nikon, NIR Apo $60/1.0$) is used for observing the particles~\cite{zu2}. A laser tweezer is built on the microscope using a cw solid-state laser operating at $1064$ nm (Aresis, Tweez $250$si) for manipulating the particles.  A charge-coupled device (CCD) video camera (iDs-UI) at a rate of 50-100 frames per second is used for video recording of the particle trajectory. A particle tracking program is used off-line to track the particles with an accuracy of $\pm 10$ nm.

\section{Results and Discussion}

\begin{figure}[!ht]
\center\includegraphics[scale=0.57]{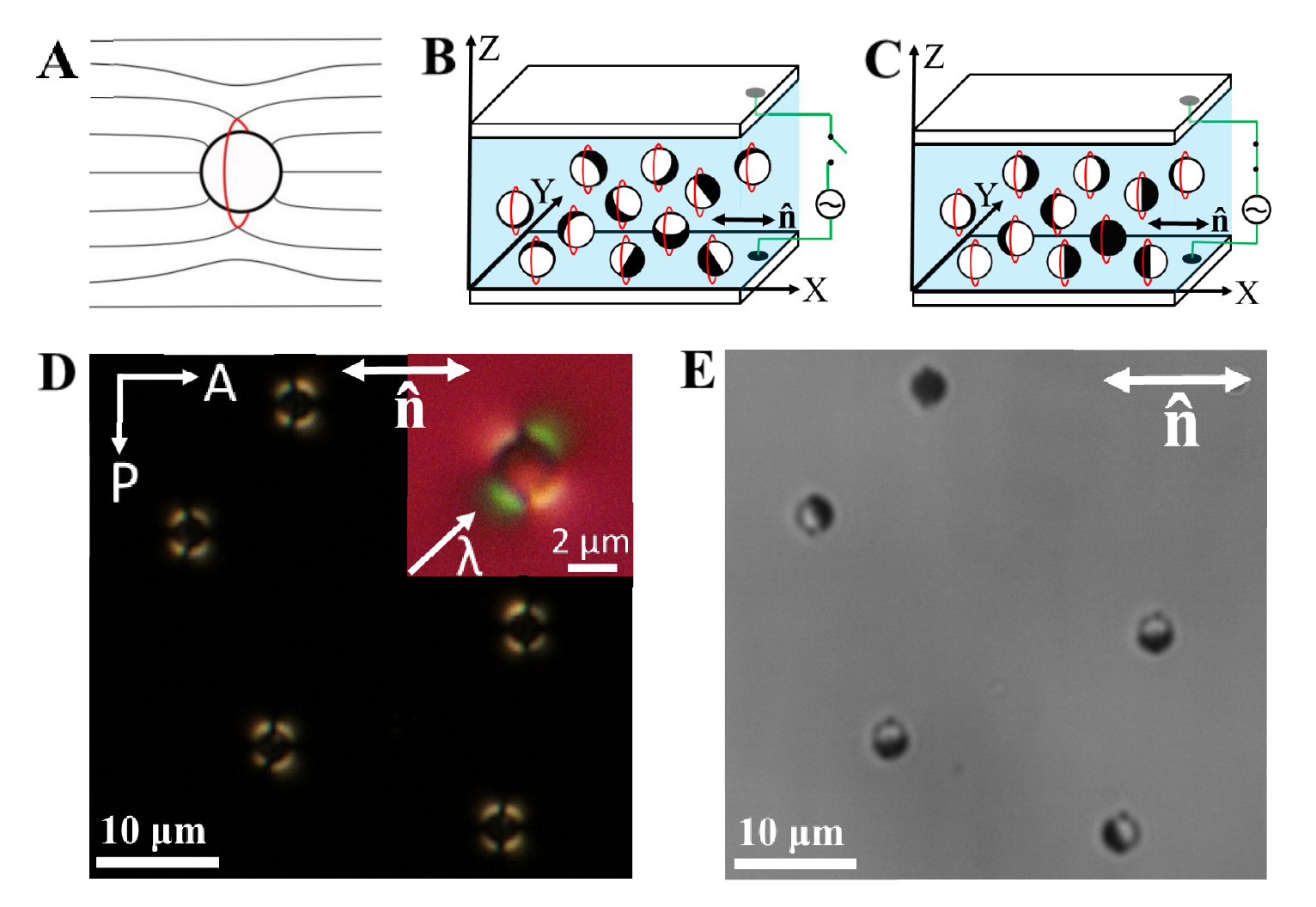}
\caption{(A) Quadrupolar elastic distortion of nematic director  ${\bf\hat{n}}$ around a spherical dielectric particle. Red circle represents Saturn-ring defect.  Janus quadrupolar particles (B) without and (C) with electric field. Note that the metal-dielectric interface becomes parallel to the field direction.  (D) Cross-polarised micrograph of a few quadrupolar particles without electric field. Inset shows the micrograph of a particle with a $\lambda$-plate (530 nm)  inserted in between the polariser and sample. (E) Image  captured by CCD camera. Dark (bright) hemisphere represents the metal (dielectric).}
\label{fig:figure2}
\end{figure}

We work in the dilute regime of the concentration (0.001wt\%) and disperse the particles in a cell whose gap is larger, but close to the diameter of the particles.  In this case the particles mostly stabilise quadrupolar director field as shown in Fig.\ref{fig:figure2}A.
 Figure \ref{fig:figure2}D shows the light-microscope texture of a few Janus particles placed between crossed polarisers. The four-lobed intensity pattern of the particles, a characteristic feature of elastic quadrupoles, is further substantiated from the texture obtained by inserting a $\lambda$-plate~(inset to Fig.\ref{fig:figure2}D). A texture of the particles (without polarisers) shows that the metal hemisphere (dark-half) of particles is oriented in different directions, always keeping the Saturn-rings perpendicular to the macroscopic director (Fig.\ref{fig:figure2}E).  Once the AC electric field is switched on along the $z$-axis, the particles reorient~\cite{sum} so that the plane of the metal-dielectric interface lies parallel to the field as shown in Fig.\ref{fig:figure2}C. Since the dielectric anisotropy of MLC-6608 is negative, the applied electric field does not influence the far field director except very close to the particles~\cite{oleg2}.

\begin{figure}[!ht]
\center\includegraphics[scale=0.46]{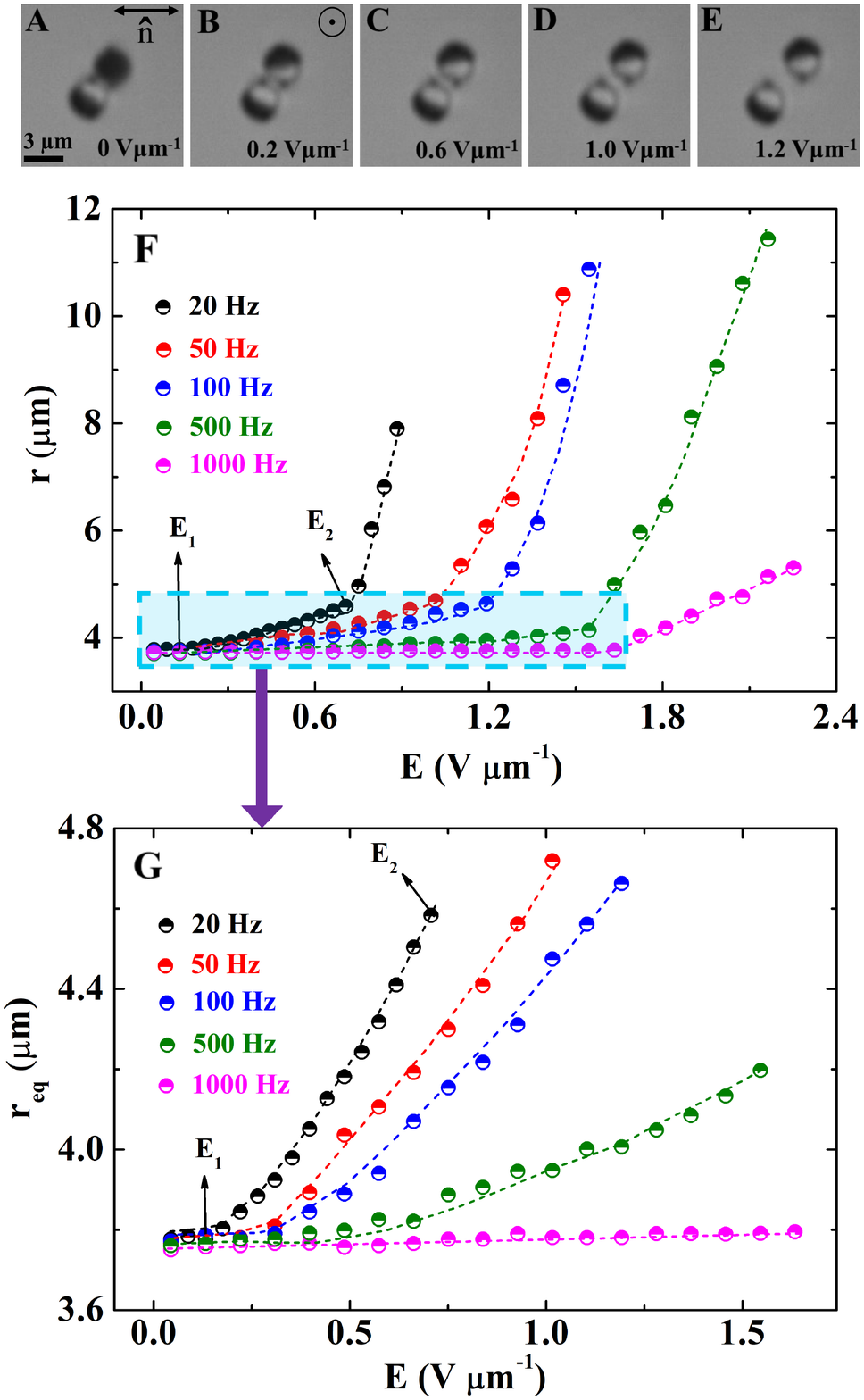}
\caption{(A-E) Effect of AC field on a pair of assembled quadrupolar Janus particles ($f$ = 100 Hz) (Movie S1, Supplemental Material~\cite{sup}). (F)  Centre-to-centre  separation $r$ with field at different frequencies. (G) Expanded rectangular region of (F) enclosed by dashed line. $E_{1}$ and $E_2$ are the first and second threshold fields, respectively and $r_{eq}$ is the equilibrium separation in the field range: $E_{1}\leq E\leq E_2$. Dotted lines are drawn as a guide to the eye. Cell thickness: 5.5 $\upmu$m.}
\label{fig:figure3}
\end{figure}

\begin{figure}[ht]
\center\includegraphics[scale=0.4]{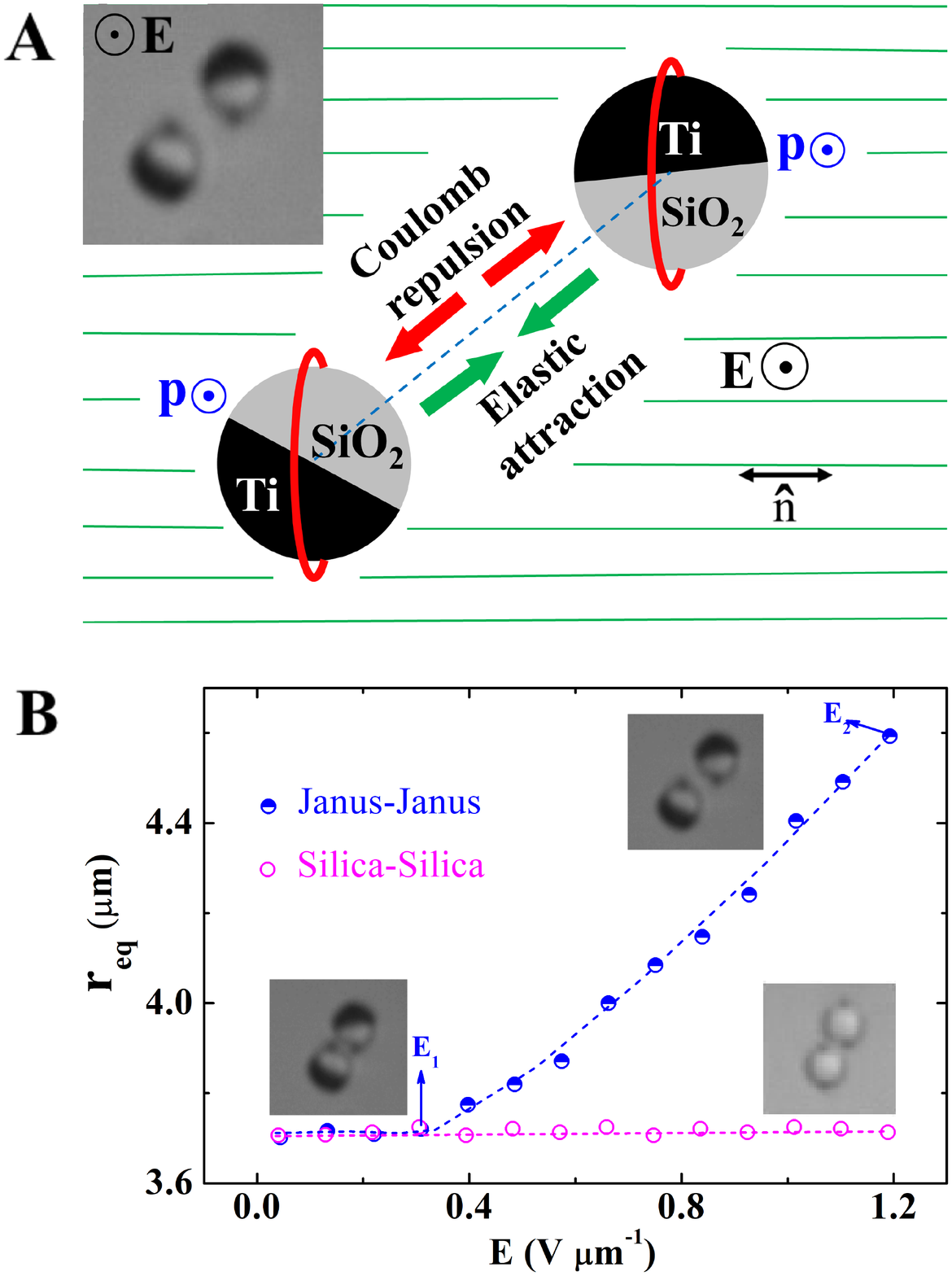}
\caption{(A) Schematic diagram demonstrating the balance of elastic and dipolar Coulomb forces. Induced dipole moments ($p$) are parallel to {\bf E} (out of plane). Green lines indicate far field director ${\bf \hat{n}}$. (Inset) CCD image of Janus particles under AC field. (B) Equilibrium separation $r_{eq}$ between a pair of Janus (blue half-filled circles) and silica (pink open circles) particles with field ($f$=100 Hz).
}
\label{fig:figure4}
\end{figure}

We focus on a pair of Janus particles which are assembled by elastic forces of the nematic liquid crystal and study the effect of field. At zero field, the line joining the centres of the particles makes an angle of 57$^{\circ}$ with respect to the director ${\bf\hat{n}}. $~\cite{ram}. At this angle the particles experience attractive elastic force. Once the electric field is switched on, the particles rotate such that the metal-dielectric interface becomes parallel to the field due to the effective induced electric dipole moment ($p_{eff}$), keeping the mutual separation unchanged as shown in Figs.\ref{fig:figure3}A and 3B (see Movie S1 in Supplemental Material, ~\cite{sup}). With increasing field amplitude, the particles are displaced in the opposite directions along the joining line of their centres as shown in Figs.\ref{fig:figure3}C to 3E. Above a certain field, the particles become free from each other's influence and swim independently. The centre-to-centre separation $r$ as a function of field at different frequencies is shown in Fig.\ref{fig:figure3}F. For a given frequency $f$, it is apparent that there is a threshold field $E_{2}$, beyond which $r$ increases rapidly. Careful observation reveals that in the low field region ($E< E_2$), there is another threshold $E_{1}$, beyond which the separation begins to increase. For clarity, the low field region of Fig.\ref{fig:figure3}F is expanded and shown in  Fig.\ref{fig:figure3}G. For example at 20 Hz, the threshold fields are $E_{1}\simeq0.2$ V$\upmu$m$^{-1}$ and $E_{2}\simeq0.7$ V$\upmu$m$^{-1}$. With increasing field the magnitude of $p_{eff}$, and consequently the repulsive dipolar force ${F}_{d}$ increases, which pushes the particles apart. At $E_{1}$, the magnitude of  ${F}_{d}$ just exceeds the elastic binding force ${F}_{el}$ and beyond $E_{2}$, the particle's state is controlled by the surrounding electroosmotic flows. In the  field range $E_1\leq E\leq E_2$, the dipolar Coulomb (repulsive) and elastic (attractive) forces are balanced  as shown schematically in Fig.\ref{fig:figure4}A, resulting  a stable system in which the equilibrium separation $(r_{eq})$ is measured. Within this range the effect of field is reversible, i.e., $r_{eq}$ reduces to zero when the field is decreased to zero.  It is observed that below $E_2$, the separation takes place along the line joining their centres which makes  an angle 57$^\circ$ with respect to the far field director ${\bf\hat{n}}$, irrespective of the orientation of the Janus vector ${\bf\hat{s}}$ (normal to the metal-dielectric interface). It indicates that the induced electroosmotic flows surrounding the particles are weak and does not affect to their equilibrium separation. Above $E_2$, the surrounding electroosmotic flows are stronger and the particles start swimming in different directions in the plane of the sample, depending on the orientation of ${\bf\hat{s}}$ as reported by Sahu \textit{et al.} recently~\cite{sd}.
%Because of Janus character, the fore-aft symmetry of the electro-osmotic flows is broken as a results of which the particles move in different directions depending of the orientation of the Janus vector~\cite{sd}. \\ 

\begin{figure}[t]
\includegraphics[scale=0.39]{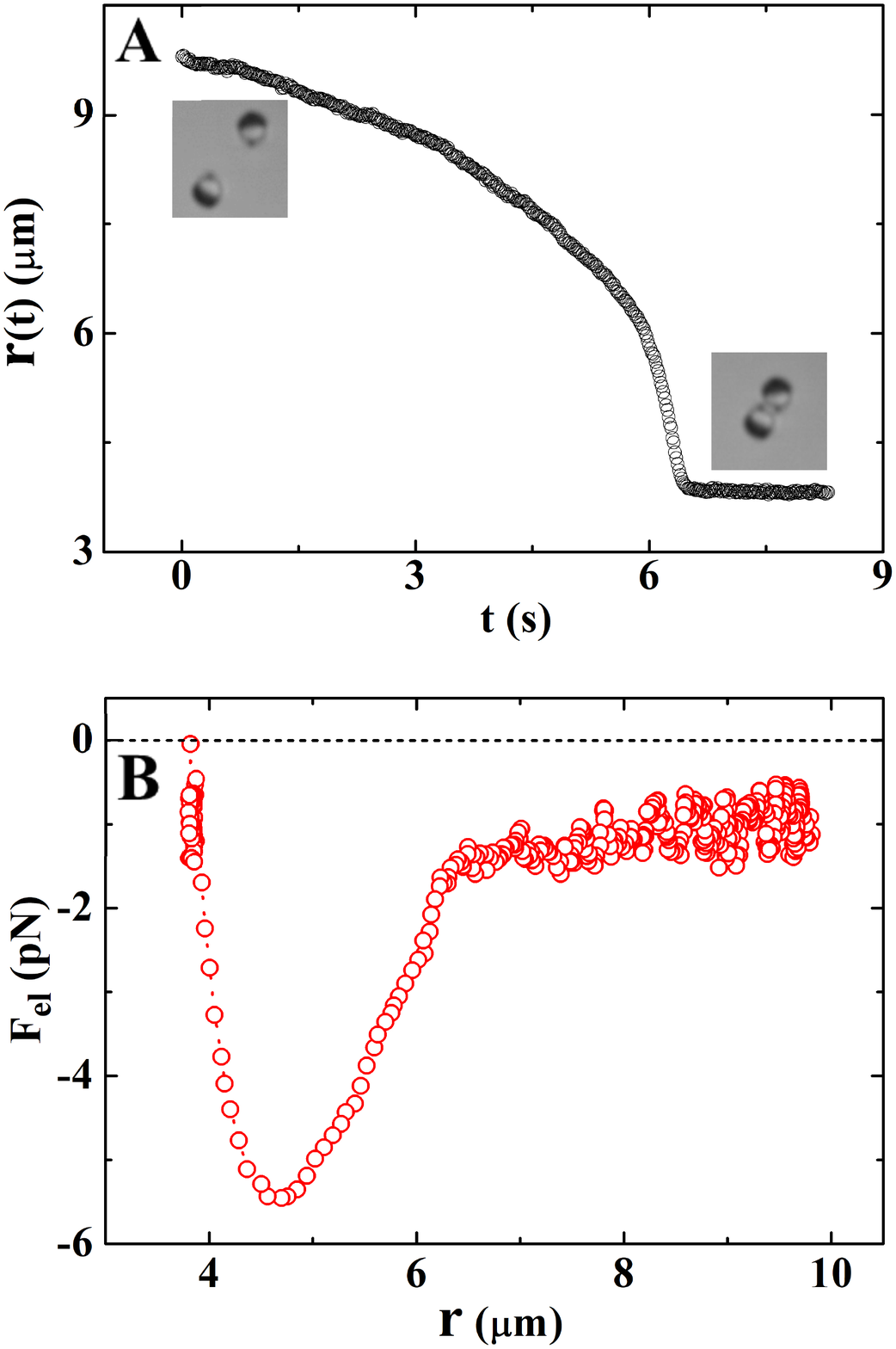}
\caption{(A) Centre-to-centre distance $r(t)$ between two Janus quadrupolar particles approaching due to elastic interaction in the absence of electric field. Inset shows the snapshots at initial ($t = 0$ s) and final time ($t$ $\simeq$ $9$ s). (B) Variation of elastic force $F_{el}$ between two Janus particles as a function of $r$ measured from the videomicroscopy.}
\label{fig:figure5}
\end{figure}

  The electric field also induces dipole moment in silica particles (non-Janus) and hence similar effect is expected to be seen in a pair of silica quadrupolar particles. To see such effects, we applied electric field and compared the results with that of the Janus particles.  Figure \ref{fig:figure4}B shows that the equilibrium separation between two silica particles does not change upto the applied field ($E\leq E_2$). Even at much higher fields the separation remains unchanged until electrohydrodynamic instability was observed.  It indicates that the magnitude of induced dipole moment in silica particles is much smaller than that of the Janus particles. This is expected as the polarisability of the silica is lower than that of the Titanium metal. As a result of which the dipolar Coulomb repulsion between the two silica particles is not sufficiently strong in the applied field range to overcome the attractive elastic forces.

\begin{figure}[ht]
\center\includegraphics[scale=0.4]{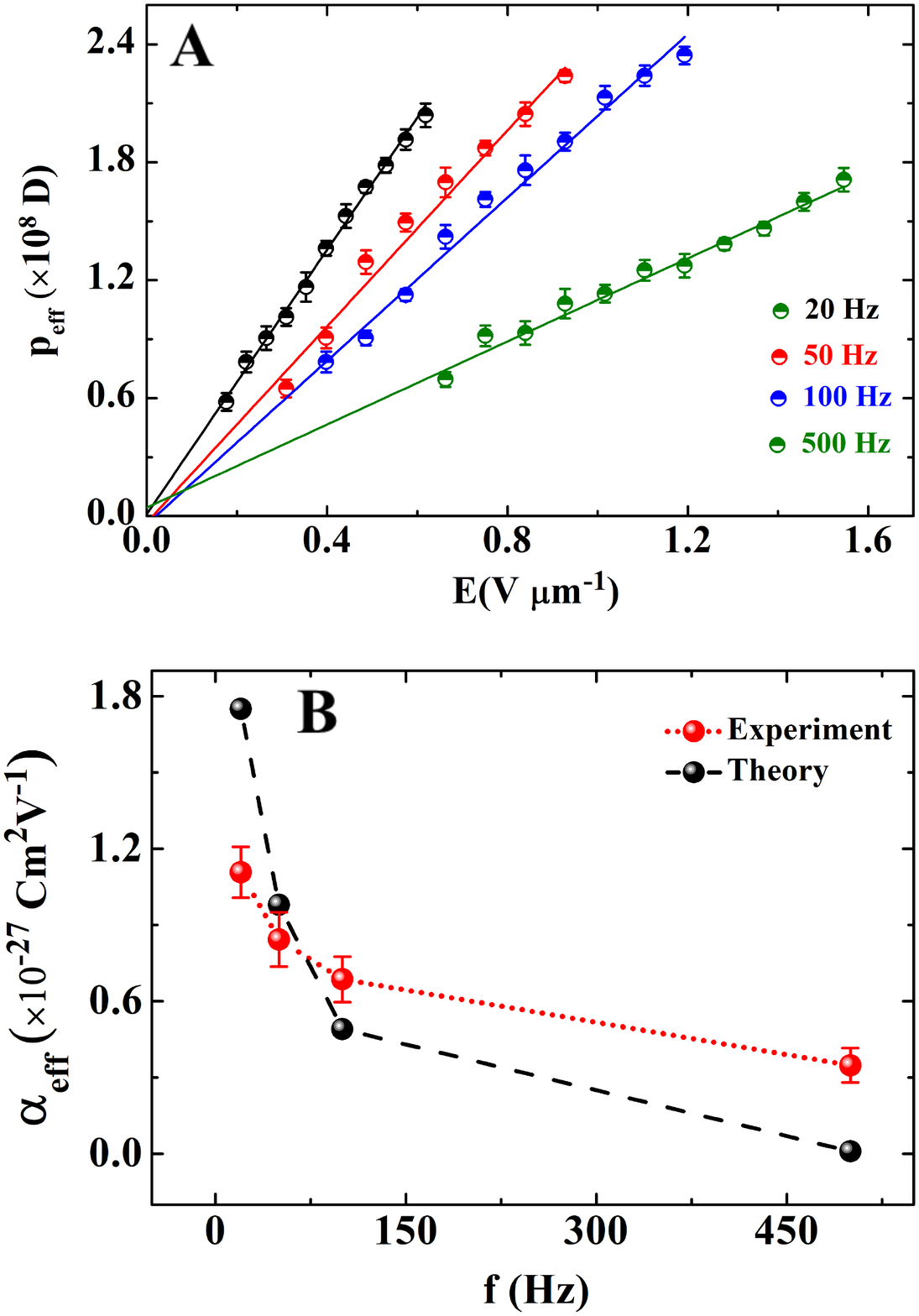}
\caption{ (A) Effective induced dipole moment $p_{eff}$ with field at different frequencies. Solid lines show the least squares fit to equation: $p_{eff} = \alpha_{eff} E$. (B) Polarisability $\alpha_{eff}$ at different frequencies obtained from  the experiments (filled red circles) and theory (filled black circles). Dotted and dashed lines are drawn as a guide to the eye. Error bars represent the standard deviation of the mean value.}
\label{fig:figure6}
\end{figure}

 The dipolar Coulomb force $F_d$ between two identical induced electric dipoles ($p_{eff}$) of the Janus particles is expressed as: 
\begin{equation}
F_{d}(r)=3p_{eff}^2(E)/4\pi\epsilon_{0}\epsilon_m r_{eq}^4(E)
\end{equation}
where $\epsilon_m$ is the relative permittivity of the medium and $r_{eq}(E)$ is the equilibrium separation. In the field range: $E_{1}\leq E\leq E_{2}$, the Coulomb force $F_{d}(r)$ is obtained by balancing it against the elastic force $F_{el}(r)$ which is obtained from the videomicroscopy in the absence of electric field. To measure $F_{el}$ initially two particles are held at a distance with the help of the laser tweezers and then allowed to interact freely, by switching off the laser. The centre-to-centre separation $r(t)$ between two particles obtained from the recorded movie is shown in Fig.\ref{fig:figure5}A. The motion of the particles in the nematic liquid crystal (NLC) is over-damped due to high viscosity. Consequently, the Stokes drag force $(F_{S})$ in a uniformly aligned NLC is in equilibrium with the elastic force $F_{el}(r)$, resulting in no acceleration. The elastic force between two particles is calculated from the numerical differentiation of the trajectory, which is given by $F_{el}(r)=-F_{S}=\zeta_{i} \frac{dr(t)}{dt}$, where $\zeta_{i}$ is the drag coefficient and the subscript, $i$ refers to the motion  either parallel to ${\bf \hat{n}}$ ($i = \parallel$) or perpendicular to ${\bf\hat{n}}$ ($i = \perp$)~\cite{d1,d2}. From the study of the Brownian motion of a free particle~\cite{ivan,igor4}, we determine the drag coefficients using the Stokes-Einstein relation $\zeta_{\parallel,\perp}=K_{B}T/D_{\parallel,\perp}$ by measuring the anisotropic diffusion coefficients $D_{\parallel}$ and $D_{\perp}$ at room temperature (see Supplemental material~\cite{sup}). In estimating $F_{el}(r)$, the average drag coefficient $\zeta = (\zeta_{\parallel} + \zeta_{\perp})/2 \approx 1.2\times 10^{-6} $ kg s$^{-1}$ is used. The variation of $F_{el}(r)$  between two Janus particles as a function of separation is shown in Fig.\ref{fig:figure5}B.  Substituting $F_{d}(r)=F_{el}(r)$  in Eq.(1), the magnitude of the effective induced dipole moment of the Janus particles can be expressed as:
\begin{equation}
p_{eff}(E)=\sqrt{(4/3)\pi\epsilon_0\epsilon_m r_{eq}^4(E) F_{el}(r)}
\end{equation}
where $\epsilon_{m}$ is the average relative permittivity of the sample  which can be written as $\epsilon_{m}=(\epsilon_{\parallel}$+2$\epsilon_{\perp})/3= 6.4$ (Supplemental material~\cite{sup}). 
The variation of calculated $p_{eff}$ using Eq.(2) as a function of field amplitude ($E_1 \leq E \leq E_2$) at different frequencies is shown in  Fig.\ref{fig:figure6}A. We observe a linear variation with field i.e., $p_{eff}=\alpha_{eff} E$, where $\alpha_{eff}$ is the effective electric polarisability of the Janus particles. It is noted that the best fit lines nearly pass through the origin. It means that the dipole moment is induced only by the electric field  and eventually it asserts that the separation ($r_{eq}$) between the two particles  (below $E_2$) occurs primarily due to the dipolar Coulomb repulsion. Figure \ref{fig:figure6}B shows $\alpha_{eff}$ obtained as a fit parameter at different frequencies. It decreases with increasing frequency as expected. 

%The polarisability obtained at different frequencies shown in Fig.\ref{fig:figure6}B decreases with frequency as expected.
 
 %It may be mentioned that during the translational Brownian motion no rotational diffusion of the particles is observed. 
%Neglecting the coupling of the nematic director to the flow field induced by the motion of the particle, one obtains $\zeta = 6\pi \eta a \approx 0.6\times 10^{-6} $ kg/s, which is in reasonable agreement with the experiment.

In what follows we theoretically calculate the polarisability of the Janus particles. For a spherical particle polarisability can be expressed as~\cite{stv2}
\begin{equation}
\alpha_{M,D}(\omega) = 4\pi\epsilon_{0}\epsilon_{m} K_{M,D}(\omega)a^{3} 
\end{equation}
where $\epsilon_{m}$  is the relative permittivity of the dispersing medium, $K(\omega)$ is the complex Clausius-Mossotti factor, $a$ is the radius of the sphere and $M$ and $D$ stand for metal and dielectric, respectively. The Clausius-Mossotti (CM) factors of the metallic and dielectric spheres are calculated by using the analytical solutions given by Ramos \textit{et al.}~\cite{ramos2} and Shilov \textit{et al.}~\cite{shi}, respectively.  The real and imaginary components of the CM factors for both metal and dielectric spheres are presented in the supplemental material~\cite{sup}. 
%: Re$[K]=(\Omega^2-1/2)/(\Omega^2+1)$ and Im$[K]=(3\Omega/2)/(\Omega^2+1)$, where $\Omega=\omega C_{D}a/2\sigma$, $a$ is radius of the particles and $\sigma _m$ is the conductivity. The double-layer capacitance $C_{D}=4\pi a^2\epsilon_m/\lambda_D$, where $\lambda_D$ is the electrical double-later thickness~\cite{ramos2}.
 The average relative permittivity of the sample obtained from the experiments is given by  $\epsilon_{m}= 6.4$ (Supplemental material~\cite{sup}).  We use superposition principle in which it is assumed that the induced dipole moment of a Janus particle is equal to the sum of the one half contributions each from a metal and a dielectric sphere so that the effective theoretical polarisability of the Janus particles can be written as $\alpha_{eff}=\frac{1}{2}(\alpha_M+\alpha_D)$. The polarisability calculated using Eq.(3) at different frequencies is shown  in Fig.\ref{fig:figure6}B, which is reasonably in good agreement with the experiments. \\

  Two further remarks are in order. First, apart from dipolar repulsion the hydrodynamic repulsion due to the overlap of electroosmotic flows may also contribute to the interparticle separation~\cite{sag1}. We present an approximate estimation of the magnitudes of the two forces. The swimming force of the particles can be  written as $F_{swim}=6\pi\eta a v$, which approximately sets the upper limit of effective hydrodynamic force. Taking radius $a=1.5$ $\upmu$m,  flow viscosity $\eta=20 $mPas and swimming velocity $v=1$ $\upmu$m s$^{-1}$ (below $E_2$), the estimated hydrodynamic force is $F_{swim}\simeq0.6$ pN. Using equation (1) and taking separation $r=4 \upmu$m, $\epsilon_{m}=6.4$ and $p_{eff}=2\times10^{8}$ D the estimated dipolar  force between two particles is given by $F_{d}\simeq7$ pN. The dipolar force is one order higher than the upper limit of the hydrodynamic force. Therefore, the effect of hydrodynamic repulsion is much smaller compared to the dipolar repulsion between the two particles. Further reasonably good agreement between the experimentally measured and theoretically calculated polarisabilities based on an approximate model  affirms that the effect of hydrodynamic repulsion on the equilibrium separation between two Janus particles below $E_2$ is negligibly small compared to the dipolar Coulomb repulsion.  Second, conceptually this method could also be employed to measure the induced dipole moments of silica particles but much higher electric field is required. Hence, appropriate liquid crystal must be chosen such that no electroconvection is observed at desired fields.   \\\\

\section{Conclusion}

In conclusion, we have studied the effect of AC electric field on a pair of quadrupolar Janus particles which are assembled in a nematic liquid crystal by the elastic forces of the director field.  The imposed electric field induces effective dipole moment in each particle which is parallel to the field direction. The dipolar Coulomb repulsion between two assembled Janus particles increases with field and beyond a particular field the repulsive force exceeds the elastic binding force of the particles as a results of which the mutual separation increases. The competing effect of these two oppositely directed forces within a certain field-range allows us to measure the effective induced dipole moments of the Janus particles which varies linearly with the field. We measured effective polarisability at different frequencies. Theoretically calculated polarisability of the particles based on the superposition principle agrees well with the experiments. Our study has focussed on spherical particles in nematic liquid crystals however, this method is applicable to all microscopic Janus particles irrespective of their shapes in variety of liquid crystals. 
%\SD{Although application of this method is limited to  liquid crystals yet it promises unexplored physical effects and applications in other anisotropic fluids.}

\section{Acknowledgments}

S.D. thanks Steve Granick for hosting his visit to IBS, UNIST which resulted in very useful discussions. S.D. also thanks Manasa Kandula and Jie Zhang for useful discussions during his stay at IBS.  We thank K.V. Raman for help in preparing Janus particles and Sriram Ramaswamy for useful discussion. This work is supported by the DST, Govt. of India (DST/SJF/PSA-02/2014-2015). S.D. acknowledges a  Swarnajayanti Fellowship and D.K.S. an INSPIRE Fellowship from the DST.


\begin{thebibliography}{99}


\bibitem{1} J. Zhang, B.A. Grzybowski, and S. Granick, Langmuir \textbf{33}, 6964 (2017). 

\bibitem{2}A. Walther and A. H. E. M{\"u}ller, Soft Matter \textbf{4}, 663 (2008).

\bibitem{stv1} J. Yan, M. Han, J. Zhang, C. Xu, E. Luijten, and S. Granick, Nat. Mater.  \textbf{15}, 1095 (2016).

\bibitem{ram}S. Ramaswamy, R. Nityananda, V. A. Raghunathan, and J. Prost, Mol. Cryst. Liq. Cryst. \textbf{288}, 175 (1996).

\bibitem{abot} Y. Gu and N. L. Abbott, Phys. Rev. Lett. \textbf{85}, 4719 (2000).

\bibitem{lub} P. Poulin, H. Stark, T. C. Lubensky, and D. A. Weitz, Science \textbf{275}, 1770 (1997).

\bibitem{stark} H. Stark, Phys. Rep. \textbf{351}, 387 (2001).

\bibitem{rev} Mu{\v{s}}evi{\v{c}} I, \textit{Liquid crystal colloids}, (Springer International Publishing AG). 2017.

\bibitem{igor1} I. Mu\v{s}evi\v{c}, Liq. Cryst. Today \textbf{19}, 2 (2010).

\bibitem{ivan1} I. I. Smalyukh, Annual Review of Condensed Matter Physics \textbf{9}, 207 (2018).

\bibitem{igor} I. Mu\v{s}evi\v{c}, M. \v{S}karabot, U. Tkalec, M. Ravnik, and S. \v{Z}umer, Science \textbf{313}, 954 (2006).

\bibitem{ivan2} Y. Yuan, Q. Liu, B. Senyuk, and I. I. Smalyukh, Nature \textbf{570}, 214 (2019).

\bibitem{ivan3} H. Mundoor, S. Park, B. Senyuk,  H. H. Wensink, and I. I. Smalyukh, Science \textbf{360}, 768 (2018).

\bibitem{oleg1} O. D. Lavrentovich, Curr. Opin. Colloid Interface Sci. \textbf{21}, 97 (2016).

\bibitem{oleg2} O. D. Lavrentovich, I. Lazo, and O. P. Pishnyak, Nature \textbf{467}, 947 (2010).

\bibitem{oleg3} I. Lazo, C. Peng, J. Xiang,  S. V. Shiyanovskii, and O. D. Lavrentovich, Nat. Commun. \textbf{5}, 5033 (2014).

\bibitem{sd} D. K. Sahu, S. Kole, S. Ramaswamy and S. Dhara, Phys. Rev. Res \textbf{2}, 032009(R) (2020).

%\bibitem{igor} I. Mu\v{s}evi\v{c}, M. \v{S}karabot, U. Tkalec, M. Ravnik, and S. \v{Z}umer, Science \textbf{313}, 954 (2006).


\bibitem{zu1} K. P. Zuhail, P.Sathyanarayana, D. Sec, S. Copar, M. \v{S}karabot, I. Mu\v{s}evi\v{c} and S. Dhara, Phys. Rev. E \textbf{91}, 030501(R) (2015)



 

\bibitem{igor2} U. Tkalec, M. Ravnik, S. \v{Z}umer, and I. Mu\v{s}evi\v{c}, Phys. Rev. Lett. \textbf{103}, 127801 (2009).

\bibitem{igor3} M. \v{S}karabot, M. Ravnik, D. Babi\v{c}, N. Osterman, I. Poberaj, S. \v{Z}umer, I. Mu\v{s}evi\v{c}, A. Nych, U. Ognysta, and V. Nazarenko, Phys. Rev. E \textbf{73}, 021705 (2006).


\bibitem{zu2} K. P. Zuhail, S. Copar, I. Mu\v{s}evi\v{c} and S. Dhara, Phys. Rev. E \textbf{92}, 052501 (2015).



\bibitem{sup} See Supplemental material for movie and supporting results.

\bibitem{sum} S. Gangwal, O. J. Cayre, M. Z. Bazant, and O. D. Velev, Phys. Rev. Lett. \textbf{100}, 058302 (2008).


\bibitem{d1} M. Rasi, R. K. Pujala and S. Dhara, Sci. Rep.,  \textbf{9}, 4652 (2019).

\bibitem{d2} V. S. Devika, R. K. Pujala and S. Dhara, Adv. Opt. Mater., \textbf{8}, 1901585 (2020).


%\bibitem{ram} S. Ramaswamy, R. Nityananda, V. A. Raghunathan, and J. Prost, Mol. Cryst. Liq. Cryst. \textbf{288}, 175 (1996).

\bibitem{ivan} C. P. Lapointe, T. G. Mason, and I. I. Smalyukh, Science \textbf{326}, 1083 (2009).

\bibitem{igor4} U. Tkalec, and I. Mu\v{s}evi\v{c}, Soft Matter \textbf{9}, 8140 (2013).



%\bibitem{ramos1} P. G. S{\'a}nchez, Y. Ren, J. J. Arcenegui,  H. Morgan, and A. Ramos, Langmuir \textbf{28}, 13861 (2012).

\bibitem{stv2} J. Zhang, J. Yan, and S. Granick, Angew. Chem. Int. Ed. \textbf{55}, 5166 (2016).



\bibitem{ramos2} A. Ramos, H. Morgan, N. G. Green, and A. Castellanos, J. Colloid Interface Sci. \textbf{217}, 420 (1999).

\bibitem{shi} V. N. Shilov,  A. V. Delgado,  F. Gonzalez-Caballero, and C. Grosse, Colloid Surface A \textbf{192}, 253 (2001).

%\bibitem{dhara} S. Dhara, and N. V. Madhusudana, J. Appl. Phys. \textbf{90}, 3483 (2001).
\bibitem{sag1} A. V. Straube, J. M. Pag{\'e}s,  P. Tierno, J. Ign{\'e}s-Mullol, and F. Sagu{\'e}s, Phys. Rev. Res. \textbf{1}, 022008(R) (2019).

\end {thebibliography}
\end{document}